# Analysing the influence of power take-off adaptability on the power extraction of dense wave energy converter arrays

Alva Bechlenberg [a], Yanji Wei [b], Bayu Jayawardhana [c], Antonis I. Vakis [a,*]

[a] Computational Mechanical and Materials Engineering, Engineering and Technology institute Groningen, University of Groningen, Nijenborgh 4, Groningen, 9747 AG, The Netherlands

[b] Eastern Institute for Advanced Study, Yongriver Institute of Technology, Ningbo, Zhejiang, 315201, China

[c] Discrete Technology and Production Automation, Engineering and Technology institute Groningen, University of Groningen, Nijenborgh 4, Groningen, 9747 AG, The Netherlands

ABSTRACT

The aim of this work is to assess the influence of different degrees of adaptability of the power take-off (PTO) system on the power absorption of dense wave energy converter (WEC) arrays. The adaptability is included in simulations through a transmission ratio that scales the force actuating the PTO relative to the force generated by the motion of a floater. A numerical model is used in which hydrodynamic interactions between floaters and nonlinearities in the PTO are considered. The lower computational cost of this numerical model makes it possible to study the power extraction of a dense WEC array in irregular waves to easily create power matrices and other performance metrics. The methodology is applied to the case study of the Ocean Grazer WEC to showcase the potential performance improvements achieved through the inclusion of a transmission ratio. The analysis shows that including a high degree of adaptability and choosing WEC array configurations and PTO designs specific to potential deployment locations early in the design process can lead to an increase in extracted power.

## 1. Introduction

To mitigate climate change, the large-scale deployment of renewable energy systems (RES) has become a top priority in decreasing carbon emissions and replacing fossil fuels altogether. Beyond wind and solar, ocean waves contain an abundant energy resource with high potential that, for the moment, remains largely untapped despite being more energy dense, reliable and predictable than wind, for example. Despite the fact that many wave energy converter (WEC) technologies have been developed and a few have even been tested in the field [1], many still require extensive research to reach full industrial scale and maturity to become competitive with, and comparable to, wind or solar energy systems.

To increase the power extraction and efficiency of WECs, R&D activities on WECs are primarily focused on optimising their shape and arranging them in arrays with different configurations [2,3], designing various power take-off systems [4] and applying control strategies to tune the floaters and improve their performance in a wider range of sea conditions [5,6]. In the paper of Garnaud and Mei [7] that investigates WEC arrays, it is concluded that dense WEC arrays broaden the bandwidth in which the device performs more efficiently. A dense WEC array is characterised by a distance between elements smaller than two times the diameter of the single floater [8]. However, when closely spaced (dense) arrays are considered, interactions between floaters become non-negligible and need to be included in simulations, which increases computational costs due to the additional complexity [9]. Correspondingly, independent of the geometry or configuration of the device, it is necessary to tune the WECs to increase their power extraction over a range of incoming wave conditions; devices can be designed to function most efficiently for a specific ocean location or incoming wave profile. In order to actively control a WEC, one or more physical components of the device need to be controllable within a range which needs to be defined when designing the device. Adaptability can be implemented through, for example, mass alteration [10,11] or tuning methods for the power take-off (PTO) system [12]. Previous research reviews control strategies acting on the hydrodynamic, PTO and grid side parameters of a WEC: at different design stages multiple parameters can be selected to be controllable to maximise the power generation of the device [13]. Similarly, in a Wave to Wire (W2 W) assessment of a WEC with hydraulic PTO, three stages of the power generation process can potentially be controlled separately: the transmission, generation and conditioning stage. Furthermore, the type of PTO system used can also influence the behaviour of the floater [14]. Hydraulic PTO systems have been used often in point absorber WECs and can be adapted with regulating valves and accumulators, while different control strategies can be applied to optimise their performance [15]. The advantages of one such hydraulic PTO with three pistons of different sizes that can be activated to create seven discrete PTO settings were investigated in previous work for both single floaters and arrays of floaters [16].

* Corresponding author.
*E-mail addresses:* a.bechlenberg@rug.nl (A. Bechlenberg), yanji.wei@eias.ac.cn (Y. Wei), b.jayawardhana@rug.nl (B. Jayawardhana), a.vakis@rug.nl (A.I. Vakis).

In this paper, a methodology is introduced in which the adaptability of the PTO system is incorporated in the simulations of the WEC power extraction in irregular sea states. In this manner, the range of adaptability that the PTO system should cover in order to maximise the power output of the WEC can be analysed before creating the physical system. The method is applied to a case study and the behaviour and influence of different application levels of the adaptability to dense WEC array performance is studied. The next section describes the methodology applied to the research, namely the numerical model and simulation parameters. Afterwards, the case study is introduced for which adaptability is analysed. Subsequently, the results are discussed to draw conclusions and recommend future research directions.

## 2. Methodology

In this section, the application of the transmission ratio is described in the general governing equations of a WEC system. Furthermore, the numerical model is described which was created in order to properly account for nonlinear forces and radiation within a dense WEC array while minimising the computational effort of the simulations.

### 2.1. Governing equations

The most general approach to simulating WEC arrays is by applying Newton's second law of motion for each WEC element and solving the resulting system of motion equations. For the purpose of this research, adaptability in the form of the parameter $\alpha$ is included in the connecting force (relative to the PTO force; see below) appearing in the equations of motion. The complete system can be described by the following second-order differential equation

$$\boldsymbol{M}_b \ddot{\boldsymbol{X}}_b = \boldsymbol{F}_e + \boldsymbol{F}_r + \boldsymbol{F}_{hs} + \boldsymbol{F}_c, \tag{1}$$

where $\boldsymbol{M}_b \in \mathbb{R}^{n\times n}$ is a diagonal matrix of masses and moments of inertia for each element with one degree of freedom (DoF), $n$ being the number of WEC elements in the array, $\boldsymbol{X}_b \in \mathbb{R}^{n\times 1}$ is the displacement vector of the WEC, and $\boldsymbol{F}_e$, $\boldsymbol{F}_r$, $\boldsymbol{F}_{hs}$ and $\boldsymbol{F}_c \in \mathbb{R}^{n\times 1}$ are the vectors of excitation, radiation, hydrostatic restoring force and the force on the cable connecting the floater and PTO, respectively. The transmission ratio $\alpha$ is defined as the ratio of the output to the input force (as shown in Fig. 2c), corresponding to the PTO ($\boldsymbol{F}_{pto}$) and cable connecting forces ($\boldsymbol{F}_c$), respectively, such that

$$\boldsymbol{F}_c = \frac{\boldsymbol{F}_{pto}}{\alpha}. \tag{2}$$

The inclusion of the transmission ratio into the motion equation of the system can be done if the dynamics of the PTO are also known (i.e., for a specific PTO design), which will be presented in Section 3.2. Only one DoF (e.g. heave) is taken into consideration, as the goal of this research is to calculate the power extraction of the WEC array rather than study the dynamics of each floater in detail. Furthermore, the mooring force is not included as it is not expected to significantly influence the DoF chosen for the power extraction of the system, i.e. heave displacement and water surface elevation are both measured relative to the same reference frame.

### 2.2. Assessment metrics

A metric commonly chosen to assess WEC designs is the power matrix, which describes the power extraction potential of the WEC system in a range of sea states defined by their significant wave height ($H_s$) and peak period ($T_p$) [17,18]. The results are depicted in matrix form and highlight the areas in which the device performs best and to what degree the power extraction differs within a certain range of sea conditions. Additionally, as presented in Babarit et al. [18], sea states with high wave height and short peak period are neglected due to their low occurrence and harsh nature in which WECs switch to survivability mode and do not extract power. The capture width ratio (CWR) (Eq. (3)) is used to further contextualise the absolute power extracted by the WEC array in each sea state. CWR is defined as the ratio between the extracted power and the wave resource over the width of the array [7,19], given by

$$CWR = \frac{P_{abs}}{BP_{avail}}, \tag{3}$$

where $P_{abs}$ is the mean power absorbed by the device in the sea state which is the output of the numerical model, $P_{avail}$ is the potential power available in one metre of wave from the sea state, and $B$ is the width of the WEC array (perpendicular to the wave direction). This dimensionless metric explains how much of the available power is absorbed by the WEC array. In this research, the available power is calculated with a simplified approximation given by

$$P_{avail} = \frac{\rho g^2}{64\pi} H_s^2 T_e, \tag{4}$$

as used in literature [20–22]. A detailed description and equations on how to calculate the available wave resource and the CWR can be found in Sinha et al. [23]. The parameter $T_e$ is the energy period that can be approximated by

$$T_e = cT_p, \tag{5}$$

with the peak period of the sea state $T_p$ and a calibration coefficient $c$ [22]. The latter is dependent on the location and its wave conditions and, thus, is defined in Section 3. The power matrix can be multiplied with the occurrence of sea states in a location, usually visualised as a scatter diagram, to calculate the annual energy production (AEP) and understand how a device performs in a specific environment, and the result can also be depicted in the same form as the power matrix. The AEP is calculated by multiplying $P_{abs}$ with time $\Delta T$ in hours per year that the sea state occurs (taken from the scatter diagram), i.e.

$$\text{AEP} = P_{abs} \Delta T. \tag{6}$$

This is usually adopted to choose the most efficient and beneficial design to be deployed for a specific location [19,24]. Babarit et al. [18], for example, follow this process to compare the performance of different devices in several locations.

### 2.3. Numerical model

Since much research has focused on simulating power extraction from waves with diverse devices, a large variety of models have been developed to approximate specific study conditions. Josh and Ronan [25], for example, analyse a number of developed models and their applicability. Time-domain models are often used to accurately represent real wave conditions and include nonlinear forces, but require higher computational resources. They are, thus, commonly used for single WECs or small WEC arrays. On the other hand, frequency-domain models exhibit reasonable accuracy with lower computational costs (since they do not include nonlinearities) and are employed for more complex devices such as large WEC arrays [26]. For the purposes of this research, a model that can analyse a dense array in irregular waves, examine interactions between floaters and include nonlinear forces is necessary. The model's main output considered is the power extracted by the WEC array. A mixed time-domain/frequency-domain (MFT) model [27] based on a harmonic balance method is used in this work. Even by including the nonlinearities of the PTO force and the system's response, this approach yields less computationally expensive simulations, making it possible to run extensive studies for irregular incoming waves and large transmission ratio ranges.

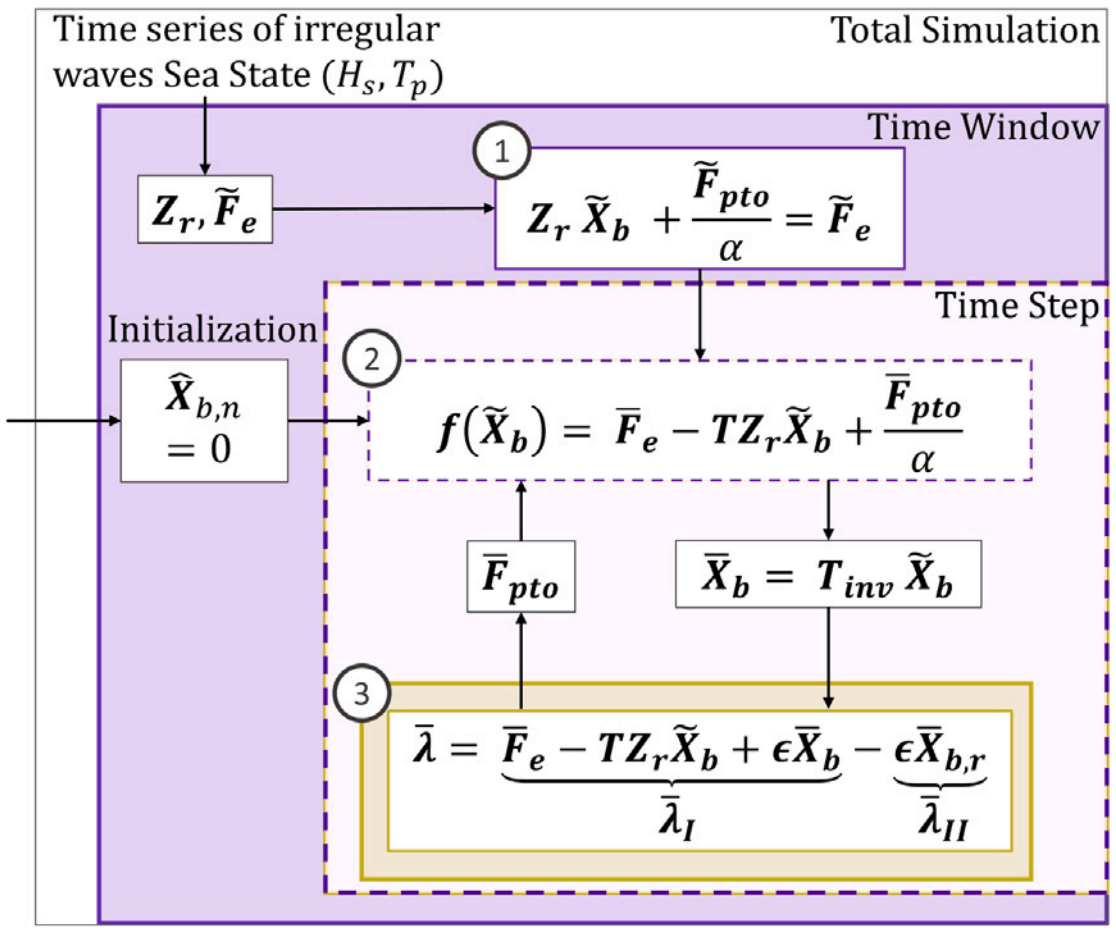

**Fig. 1.** Flowchart describing the procedure of the MFT numerical method.

The overall numerical procedure of the MFT model is depicted in Fig. 1. Each sea state is computed separately: a time series of irregular waves is generated for the total simulation time with the parameters $T_p$ and $H_s$ and divided into time windows ($tw$); this is the pre-processing of the model. On the basis of the pre-processing input, for each buoy $b$ in the WEC array of $N_b$ floaters, the excitation force in the frequency domain $\tilde{F}_e$ and the condensed dynamic stiffness $\boldsymbol{Z}_r$ are generated for each time window (as shown in purple in Fig. 1). The parameter $\boldsymbol{Z}_r$ is a combination of the stiffness of the floater and the PTO system (in case this influences the motion of the floater). The array's system of equations of motion are generated in matrix form to include the harmonic motion equation of each floater in the frequency domain (Step 1 in Fig. 1). Due to the unknown nature of the Fourier coefficient of the displacement for each floater $\tilde{\boldsymbol{X}}_b$, it is not easy to solve the system of equations as the PTO force $\tilde{\boldsymbol{F}}_{pto}$ is undetermined. Furthermore, it is not realistic to solve for $\tilde{\boldsymbol{F}}_{pto}$ in the frequency domain as it can contain nonlinearities. Instead, each time window is divided into equally spaced time steps in which the equations of motion are solved with the MFT method, as seen in Step 2 of Fig. 1. The time steps are applied to include the non-periodicity of the incident wave. Each time step in the analysed time window is computed in parallel and the floater displacement $\hat{\boldsymbol{X}}_b$, approximated by a truncated Fourier series, is initialised as 0 at the beginning of each time step. The PTO and excitation forces are described in the time domain (shown as a yellow section in Fig. 1), while the buoy displacement is transformed back and forth between the time and frequency domains through the application of (inverse) discrete Fourier transform matrices ($\boldsymbol{T}, \boldsymbol{T}_{inv}$). The PTO force is computed in the time domain $\bar{\boldsymbol{F}}_{pto}$ (Step 3 in Fig. 1) by applying Lagrange multipliers $\bar{\lambda}_I$, $\bar{\lambda}_{II}$ and $\epsilon$ as a penalty coefficient. $\bar{\lambda}_{II}$ is an additional term that depends on the PTO system type (e.g. hydraulic) and is not always necessary. An iterative process in each time window is applied until the system residual in the MFT model (Step 2 in Fig. 1) is smaller than a predefined target tolerance. In case the residual is larger, the buoy displacement $\tilde{\boldsymbol{X}}_b$ is revised and the next iteration is initiated. The governing equations and a more detailed explanation of the method can be found in Wei et al. [27].

The outputs of the simulations are the power extracted at each time window in addition to the displacement, velocity and force of each floater. The results for each sea state and each transmission ratio yield a 4-dimensional data-set ($H_s$, $T_p$, $tw$, $\alpha$) from which the maximum power extracted per sea state or per time window is obtained and the effect of different degrees of adaptability of the transmission ratio can be analysed. We should clarify that, in the present work, we do not focus on active control of the WEC array, but on the assessment of the best case scenario of adaptability in order to further include this parameter in the design process of the PTO system for a specific dense WEC array design (see Section 3). In the simulations performed for this research, no limitations such as slamming, stroke or force restrictions were implemented, and only the transmission ratio adaptability was delimited. These limitations were investigated and identified to be relevant in previous work such as De Backer et al. [10] and Sinha et al. [23].

### *2.4. Levels of adaptability*

PTO adaptability can be achieved in various ways, as discussed previously. Before discussing potential implementations in (physical) models of specific PTO designs, we will first assess the potential of PTO adaptability to argue why and to what extent it is worth including this feature into a physical system. To give a beneficial overview of the increase in power extraction through the use of an adaptable transmission ratio, we defined four separate cases with increasing adaptability. It is assumed that a higher level of adaptability leads to higher costs in the design and operation of the system; thus, it is desired to find the appropriate strategy to include a sufficiently high level of adaptability with the least amount of resources necessary. Although the transmission ratio of each floater-PTO element in the array can be tuned separately in the envisioned system, for the purposes of this research, the transmission ratio is adapted uniformly over the entire WEC array. Referring to Eq. (2), the easiest way to conceptualise the role of the transmission ratio in controlling adaptability would be to imagine a lever connecting the floater and PTO sides of the cable, whose fulcrum can be adjusted to achieve mechanical advantage. For less energetic waves (i.e. a smaller cable force $\boldsymbol{F}_c$), a larger value of $\alpha$ would correspond to a fulcrum position that increases the PTO force ($\boldsymbol{F}_{pto}$), thereby enabling the PTO to extracted energy from low-energy waves. The opposite would hold for small values of $\alpha$. The four degrees of adaptability are defined as follows:

1. No adaptability: This is simulated to put the advantages of adaptability into context and to give a reference case. In the first scenario, the transmission ratio is neglected ($\alpha = 1.0$) and the WEC array's performance without adaptability is presented.
2. Constant transmission ratio for each Sea State (Slow Tuning [28]): In the second case, the transmission ratio is limited to be uniform for the entirety of the sea state; one value for $\alpha$ for the entire length of the simulation is applied and the value of $\alpha$ that gives the maximum power extracted is selected.
3. Restricted adaptability: The third scenario increases the adaptability by permitting the transmission ratio to change for each time window with a limitation to the extent of change from one time window to the next. The initial value of $\alpha$ is assumed to be 1.0 for the first time window of each sea state simulation, i.e. $\alpha(tw = 1) = 1.0$; for subsequent time windows, the transmission ratio that maximises the power extraction of that time window within the range $\alpha_{tw} = \alpha_{tw-1} \pm 0.2$ is chosen and the constraint that the ratio needs to remain within the initially defined range of 0.1 to 2.0 is added. Restricted adaptability would be used to avoid sudden force changes and ensure the survivability and durability of the physical components.
4. Adaptability without limitations (Fast Tuning or Wave-to-Wave Tuning [28]): In this last case, the transmission ratio can be tuned for each time window without any limitations. Specifically, the transmission ratio generating the highest power is chosen for each time window. It should be noted that achieving wave-to-wave tuning may not be possible for any PTO.

In Section 3, a hydraulic PTO is introduced and used as a case study. This PTO can adapt to one of 7 discrete settings and is therefore suitable for wave-to-wave tuning (case 4). The results and insights gained from the optimisation of the transmission ratio can provide a starting point in selecting the most suitable dimensions and adaptability settings of the physical PTO system based on a chosen deployment location.

## 3. Case study

In order to assess the effect of a variable transmission ratio on PTO adaptability for a dense WEC array, defining the design of the latter is necessary as hydrodynamic interactions between floater elements dominate the array's response to different sea states. Hence, we introduce the Ocean Grazer WEC (OG-WEC) as a case study of a dense WEC array [29] and analyse its deployment at the Ekofisk location for which we have open access to wave data [30].

### 3.1. The Ocean Grazer WEC (OG-WEC)

The Ocean Grazer (OG) is a hybrid RES that combines the extraction of energy from wind and wave resources with on-site storage. The system is composed of a wind turbine placed on a central pillar surrounded by a dense WEC array (the OG-WEC) [29]. The submerged bottom-fixed base houses the hydraulic PTO system, which is connected to each floater on the ocean surface through a cable and an adaptable transmission, and the storage system that stores the working fluid in an inflatable bladder around the base structure that is compressed by the hydrostatic pressure of the surrounding ocean water (Fig. 2c). The potential energy stored in the bladder can be used to generate electricity on demand by allowing the working fluid to flow back into the rigid reservoir (at atmospheric pressure) through hydro-turbines. Since the pressure difference (hydraulic head) between the working fluid in the bladder and that in the rigid reservoir is linearly dependent on the depth at which the gravity-based foundation is installed, the storage capacity for the same working fluid volume scales linearly with depth. In turn, this hydraulic head affects the mass of the PTO system, which is why the depth of deployment is also relevant to the performance of the PTO.

The WEC array of the so-called OG3.0 design is depicted in more detail in Fig. 2a–b: 18 floaters with a conical shape are arranged in a honeycomb structure to maintain the same distance between floaters, yielding an omnidirectional WEC array that can work independent of the dominant wave propagation direction. In the OG3.0 design, the central pillar is meant to be supporting an offshore wind turbine and is also used to ventilate the rigid reservoir such that its pressure is always atmospheric. Nevertheless, other hybrid device configurations such as wave-storage are possible where the central pillar is not necessary and maintaining the atmospheric pressure in the rigid reservoir may be achieved through other means (e.g. via an umbilical to the surface). For the purposes of this research, we account for the presence of the central pillar and incorporate its radiation effects on the floaters [31].

The PTO system coupled to each floater comprises a piston moving within a pipe that connects the internal reservoir to the external flexible bladder (Fig. 2c). The PTOs extract energy by using upstrokes to push the working fluid into the bladder, introducing nonlinear forces to the system [27,29]. In connection with the design of the OG-WEC, a physical PTO system called the multi-pump, multi-piston PTO (MP$^2$PTO) was developed, modelled and validated experimentally; this comprises three separate pistons with different diameters connected to each floater [32]. Depending on the incoming wave, different combinations of the three pistons can be activated adapting the system in order to maximise the power extracted [16]. This specific PTO system is a physical approximation of a restricted transmission ratio range with 7 adaptability steps (all possible combinations of three pistons). The PTO dimensions and transmission ratio range are chosen based on previous research using frequency domain simulations [31] and the PTO efficiency is adopted from previous research with experimentally validated time domain simulations [32]. Limiting the transmission ratio to a range between 0.1 and 2 is a parameter constraint chosen to ensure that the proposed cases of adaptability can be realised in a PTO system, considering the associated costs of adaptability and controllability: a PTO could be made increasingly adaptable at the expense of complexity and cost. Ideally, the WEC array would maximise the power extraction in the most frequently occurring and energetic sea states with minimum need for adaptability ($\alpha = 1.0$). Sea states with higher and lower energy can be harvested by utilising the entire range of $\alpha$.

**Table 1**
Summary of parameters necessary for the simulation.

| | Description | Symbol | Quantity | Unit |
|---|---|---|---|---|
| WEC array | Diameter of Floater | $d_b$ | 5 | m |
| | Draft of Floater | $D_b$ | 3 | m |
| | Buoy Mass | $m_b$ | 15000 | kg |
| | Number of Floaters | $N_b$ | 18 | – |
| | Width of Array | $B$ | 52 | m |
| PTO | Transmission Ratio Range | $\alpha$ | [0.1–2] | – |
| | Piston Mass | $m_p$ | 1000 | kg |
| | Piston Diameter | $d_p$ | 1 | m |
| | Piping Length | $l_p$ | 30 | m |
| | PTO Efficiency | $\eta_p$ | 0.9 | – |
| Location | Water Depth | $D$ | 60 | m |
| | Coordinates | | 56.54N; 3.22E | ° |
| | Calibration Coefficient | $c$ | 0.9 | – |
| | Peak Period Range | $T_p$ | [3–14] | s |
| | Significant Wave Height Range | $H_s$ | [0.5–8] | m |
| Simulation | Total Simulation Time | $t_{tot}$ | 5000 | s |
| | Time Window Length | $tw$ | $2T_p$ | s |

All parameters necessary for the simulation and their values are given in Table 1.

### 3.2. Applied numerical model

To apply the governing equations as described in Section 2, the nonlinear PTO force is modelled using a detailed description of the hydraulic PTO system connected to each floater of the WEC array of the case study. The hydraulic PTO moves in the heave direction and only extracts energy in the upstroke motion such that the nonlinear pumping force acting on the PTO's piston(s) can be described as

$$F_p = \begin{cases} \left[\rho g \left(D - h_r\right) + \rho l_p \ddot{z}_p + \rho \dot{z}_p^2\right] A_c & \text{for } \dot{z}_p > 0 \\ 0 & \text{for } \dot{z}_p \le 0, \end{cases} \tag{7}$$

where $z_p$, $\dot{z}_p$ and $\ddot{z}_p$ are the piston kinematics in heave. The pumping force is described through three terms, all of which include the piston area $A_c$. It should be noted that, in the physical realisation of this PTO system [32], each of the three pistons has a different area and adaptability can be included through an effective piston area: the sum of the areas of activated pistons. In the present work, where we include adaptability via a transmission ratio, the piston area is taken to be constant ($A_c = \pi d_p^2$) and the pumping force is scaled with the transmission ratio, as will be explained below. The first term in the pumping force is the hydrostatic force from the hydraulic head calculated as a function of the working fluid density $\rho$, the gravity $g$, the water depth $D$ and the depth of the reservoir $h_r$ (as represented in Fig. 2c). The second term describes the inertia of the working fluid within the system which is the product of the density $\rho$ and piping length $l_p$ with the piston acceleration $\ddot{z}_p$, while the third term represents the dynamic pressure of the working fluid. When the velocity of the piston is positive ($\dot{z}_p > 0$), the PTO system is in upstroke motion while, if $\dot{z}_p \le 0$, the pumping force is zero and the check valves of the PTO system ensure there is no backflow and no power is lost.

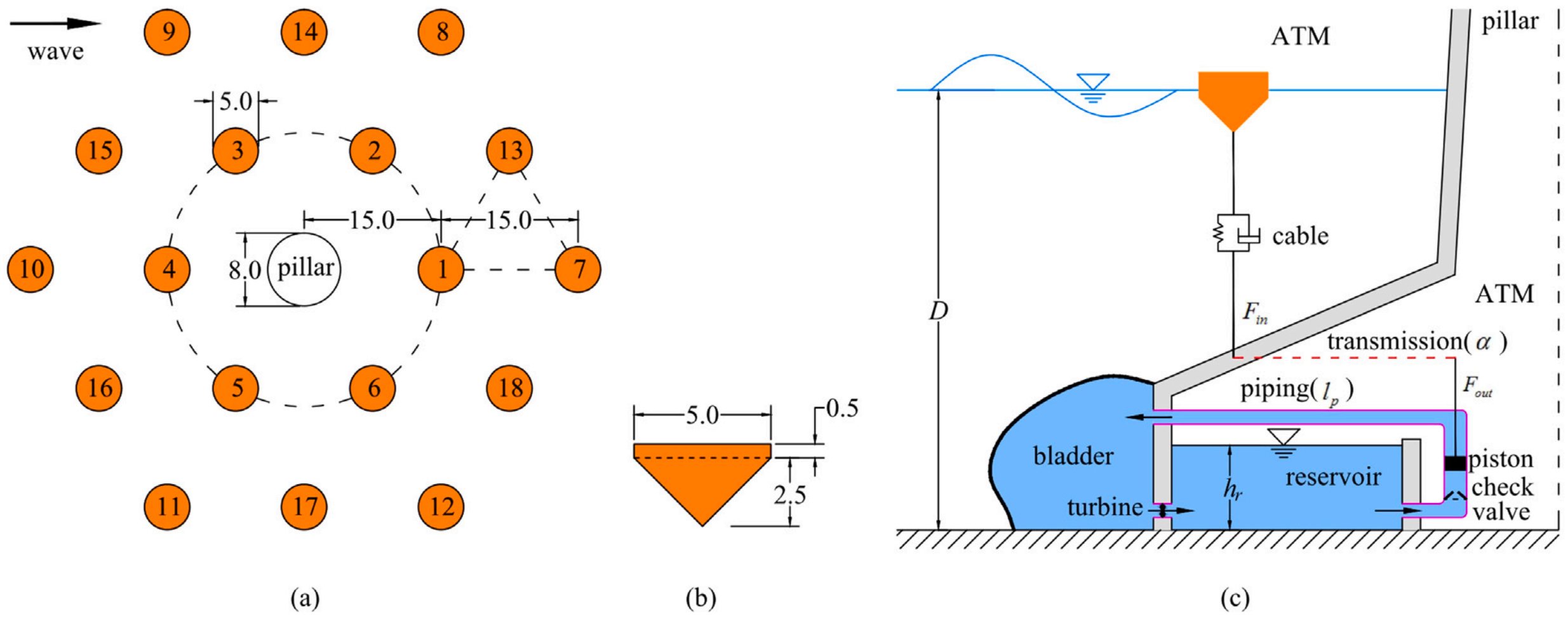

**Fig. 2.** (a) OG3.0 array configuration with central pillar and (b) the shape of each floater within the array; (c) internal System of the OG3.0 with PTOs, transmission ratios and storage subsystem. Reprinted from [29].

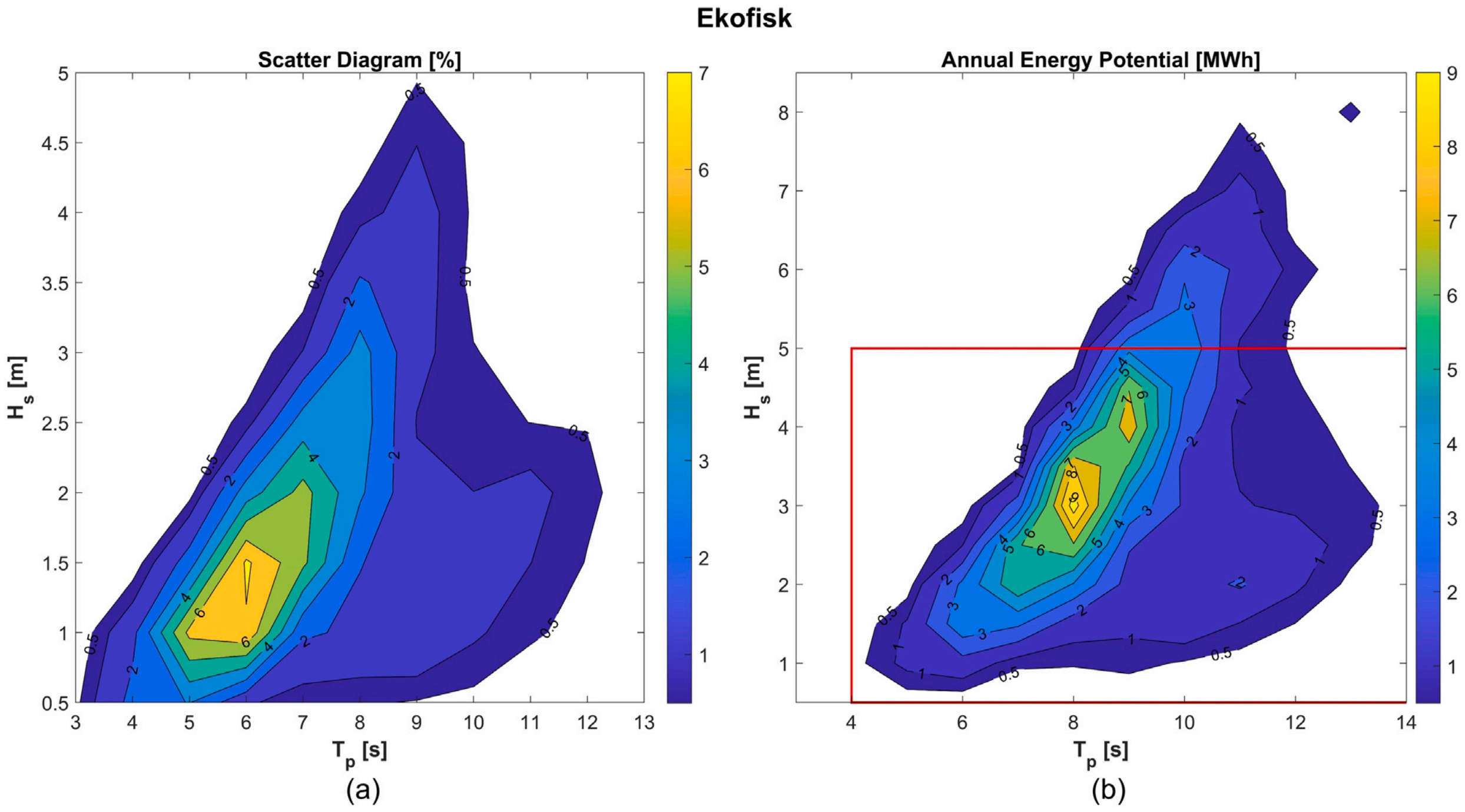

**Fig. 3.** For Location Ekofisk in the North Sea: (a) scatter diagram of % of time per year one sea state occurs (>0.5%); (b) annual energy potential of each sea state (>1 kWh).

The equation of motion of the hydraulic PTO derived in Wei et al. [27] assumes that no power is lost in the transmission system (i.e., $P_{pto} = P_c$) such that $\ddot{z}_p = \ddot{X}_b/\alpha$ based on Eq. (2) and

$$m_p \frac{\ddot{X}_b}{\alpha} = -F_p - \alpha F_c, \tag{8}$$

where $m_p$ is the mass of the piston. Solving Eq. (8) for the cable force $F_c$ and substituting that into Eq. (1) yields the following equation for a single WEC array element with an adaptable nonlinear hydraulic PTO (one OG-WEC unit)

$$\left(M_b + \frac{m_p}{\alpha^2}\right)\ddot{X}_b = F_e + F_r + F_{hs} - \frac{F_p}{\alpha}. \tag{9}$$

Due to the influence of the PTO system on the buoy's behaviour, $Z_r$ includes the stiffness of the PTO system and $\bar{\lambda}_{II}$ has to be included in the time domain simulation of the numerical method as described in Section 2.3.

### 3.3. Location

On the basis of Beels et al. [30], the location Ekofisk in the North Sea is chosen to analyse the behaviour of the OG-WEC. Ekofisk is an oil platform in the far offshore Norwegian section of the North Sea with 60 m depth. The hourly significant wave height ($H_s$) and peak period ($T_p$) between 2010 and 2020 for the region of the coordinates of Ekofisk were downloaded from the ERA5 database [33]. This database has been established by the European Centre for Medium-Range Weather Forecasts (ECMWF) and includes hourly estimates of climate variables (for this research, oceanic) covering the Earth with a 30 km resolution. It shows hourly global estimates based on historic data and advanced modelling systems [34]. On the basis of the downloaded parameters, the scatter diagram and annual energy potential (Fig. 3) were generated. The scatter diagram (Fig. 3a) shows the range of probabilities of occurrence from 0.5% to the highest of 7.2%. The annual energy potential (Fig. 3b) shows a wider range when considering all sea states that produce more than 0.5 MWh per year. The red area in Fig. 3b delimits the range of sea states which includes 90% of occurrences. The JONSWAP spectrum was selected for the simulations of the power absorption of the WEC array, which is suitable for the North Sea [23, 35]. Furthermore, to calculate the power and energy available at this location as described in Eqs. (4) and (5), the calibration coefficient $c$ is chosen to be 0.9 in accordance to the JONSWAP spectrum [22]. For the purpose of this research and in line with the range of sea states of the location, the results of the power extraction of the WEC array are shown within the ranges of $H_s = [0.5–8]$ m and $T_p = [3–14]$ s.

**Table 2**
Summary of analysed Cases.

| Case | Description | Explanation |
|---|---|---|
| Case 1 | $\alpha_{tw} = 1$ | $\alpha$ is constant over all sea states and is equal to 1; thus, this case represents no adaptability |
| Case 2 | $\alpha_{tw} = \alpha_{P_{max,SS}}$ | $\alpha$ is chosen to maximise the power extracted over the entire sea state and is kept constant within the sea state simulation |
| Case 3 | $\alpha_1 = \alpha_{P_{max,1}}$, $\alpha_{tw} = \alpha_{tw-1} \pm 0.2$ | For each time window, $\alpha$ is chosen to maximise the power extracted in the time window within the range of $\alpha_{tw-1} \pm 0.2$ |
| Case 4 | $\alpha_{tw} = \alpha_{P_{max,tw}}$ | For each time window, $\alpha$ is chosen to maximise the power extracted in the time window without limitations |

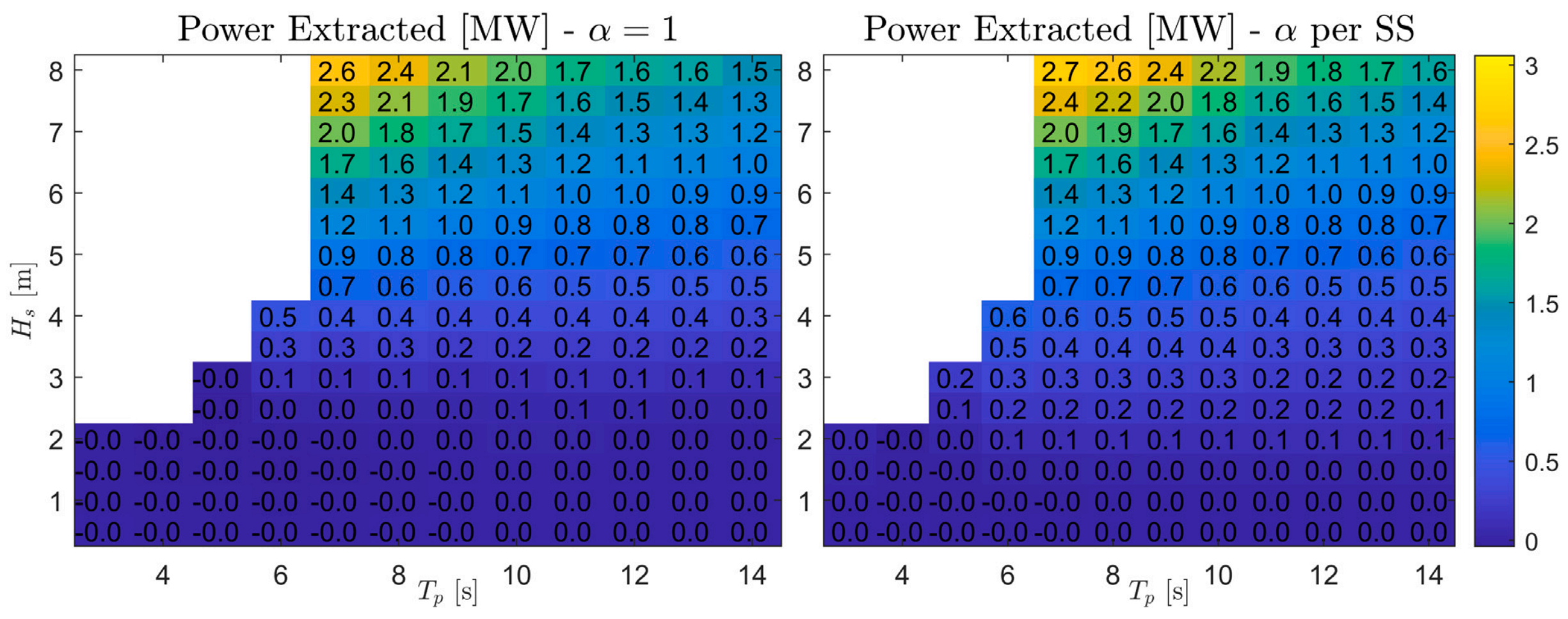

**Fig. 4.** Total power absorbed: (a) Case 1, without adaptability ($\alpha = 1$); (b) Case 2, uniform transmission ratio optimised for the entire sea state.

## 4. Results

Four cases are defined in which the degrees of adaptability described in Section 2.4 and summarised in Table 2 are applied to the WEC array introduced in the case study (Section 3). The power matrices, difference in total power absorbed, capture width ratio (CWR) and the annual energy production (AEP) are used to assess the influence of the transmission ratio $\alpha$ on the WEC array's performance.

### *4.1. Power matrices*

For Cases 1 and 2 the power matrices were generated and shown in Fig. 4. In addition to the power matrices of Cases 3 and 4 (Fig. 5a-b), the performance of these is described by plotting the absolute difference in power extracted in comparison to Case 2 in a power matrix format (Fig. 6). For these cases, the total power output achieved in Case 2 is subtracted from the total power output of Cases 3 and 4 to show the MW of difference in output as a result of increasing the level of adaptability. This is meant to highlight the improvements or changes to the performance instead of analysing the total power extracted.

In Case 1 (Fig. 4a), the maximum power output achieved within the defined sea state range is 2.6 MW; the power increases significantly due to $H_s$ and moderately due to $T_p$, showing a linear increase in power extraction. With increasing $H_s$, an increase in influence of the peak period on the power output can be identified. At low peak periods and wave heights (3–9 s $T_p$ and 0.5–2 m $H_s$), the average power output for the sea states predicted by the model is negative (dark blue area), reaching values as low as −37.7 kW at $H_s = 2$ m and $T_p = 4$ s. This can be attributed to the numerical error introduced by the Gibbs phenomenon in the harmonic-balance-based model when the heave motion in the PTO is too slow (low velocities and large forces) [27,32]. In Wei et al. [27] Fig. 3, the time series of floater motion, piston force and power extraction show the possibility of negative power and force.

For Case 2, the power matrix resembles that of Case 1 (Fig. 4b), suggesting that the level of adaptability in this case is not sufficient to hinder negative power output in low $T_p$ and $H_s$ (maximum of −20.9 kW at $H_s = 2$ m and $T_p = 4$ s) but is able to decrease negative occurrences in terms of both magnitude and occurrence. The maximum power extracted is at high $H_s$ and $T_p$ and the performance is similar to that of Case 1; the main difference in power output can be found at low $H_s$ and $T_p$.

Cases 3 and 4 show slight differences in power extraction behaviour that can be understood when analysing Fig. 6. Although the power matrices in Fig. 5 show increases in power extracted but suggest the same linear behaviour as Cases 1 and 2, focusing on the difference in absolute power shows changes at certain peak periods. Similarly to Cases 1 and 2, these cases show maximum power extraction at high $H_s$ and low $T_p$, and negative results at low $H_s$ and $T_p$. In Case 3, negative values are seen in eleven sea states within the range from $H_s = 0.5$ m, $T_p = 7$ s to $H_s = 2$ m, $T_p = 4$ s with a minimum of −9 kW. For Case 4 negative values are found in seven sea states between $H_s = 0.5$–1.5 m and $T_p = 4$–6 s with a minimum of −1.9 kW. In both cases, sea states below $H_s = 3.5$ m extract less than 0.5 MW and their maximum power extraction is at the same sea state, whereas Case 3 achieves 2.9 MW, while an increase in power extraction of approximately 6% can be observed for Case 4, for which the maximum power extraction reaches 3.1 MW. These results suggest that more adaptability (wave-to-wave tuning) results in greater power extraction.

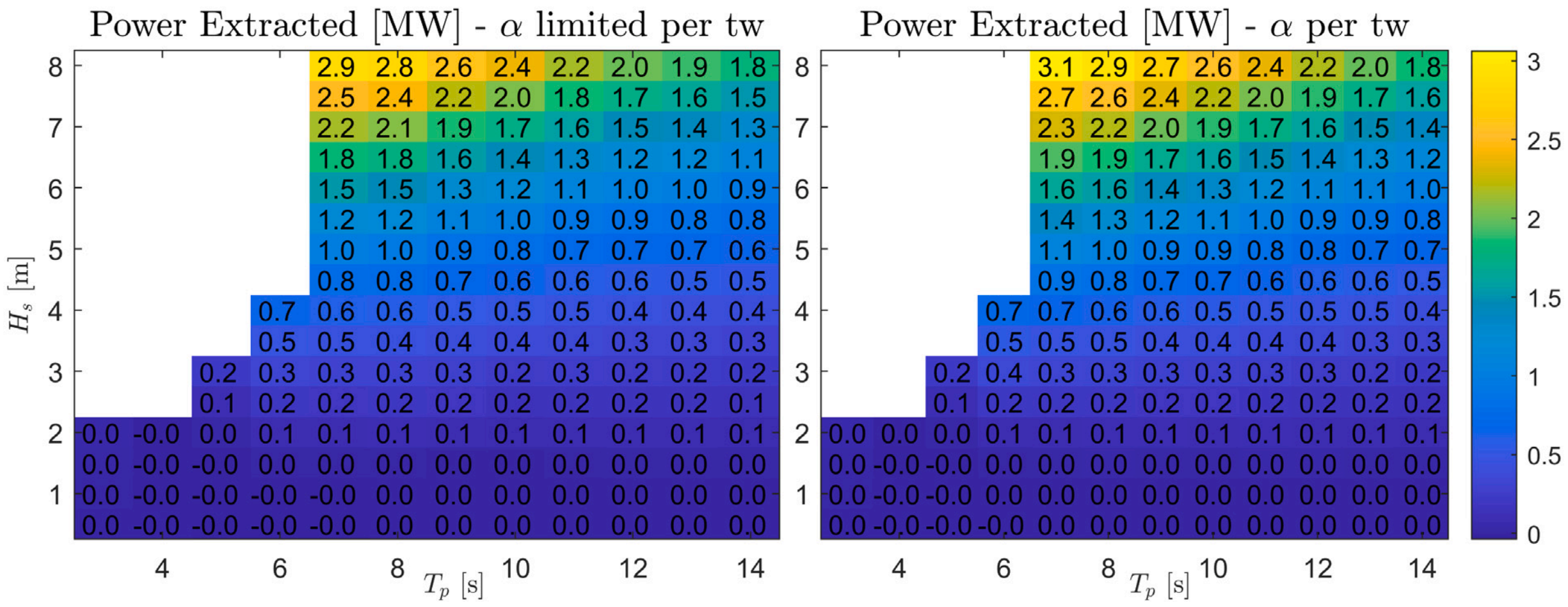

**Fig. 5.** Total power absorbed: (a) Case 3, transmission ratio adapted for the time window restricted to (±0.2); (b) Case 4, transmission ratio adapted for the time window without restriction.

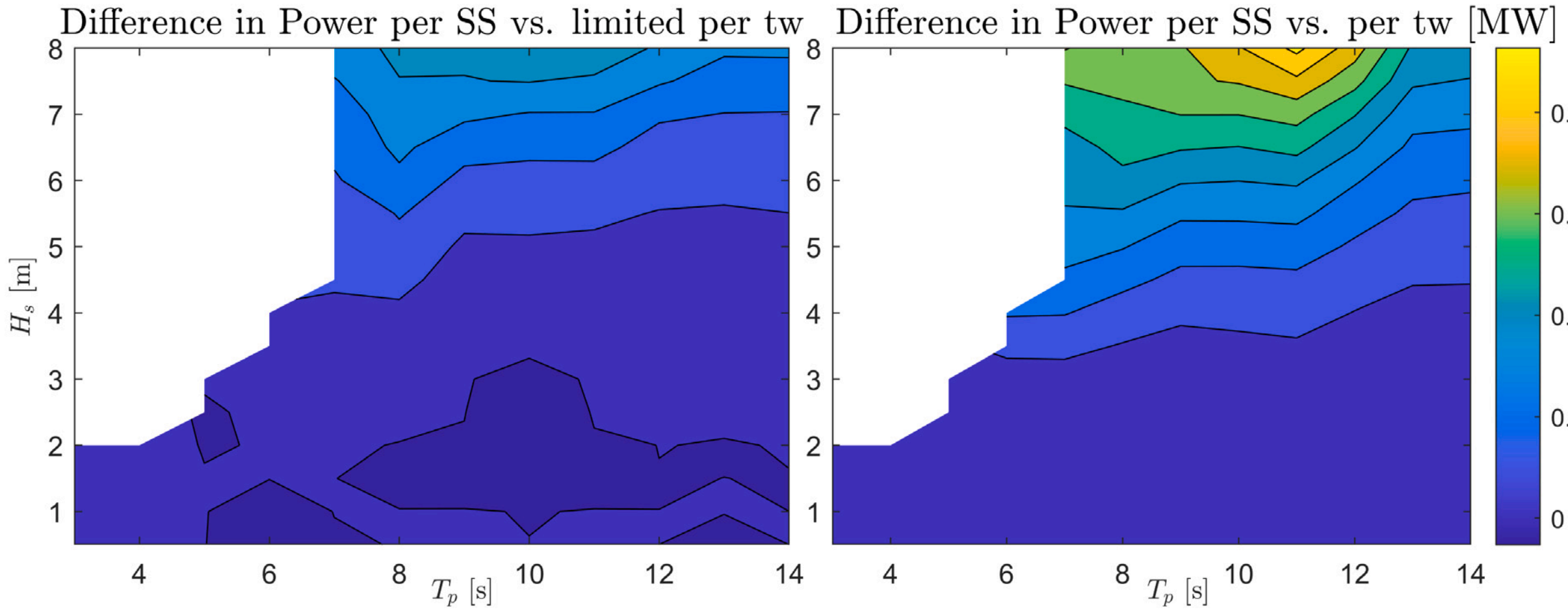

**Fig. 6.** Absolute power difference between Case 2 and: (a) Case 3; (b) Case 4.

When analysing the increase in power extraction by adopting a higher adaptability level (Cases 3 and 4) and comparing the results to Case 2 (Fig. 6), Case 3 is found to result in less than 0.25 MW of improvement (referring to the maximum change observed at sea states with the largest $H_s$); at sea states with moderate $H_s$ (0.5–5.5 m) – and especially at higher peak periods – this adaptability strategy even leads to a small decrease in power output compared to Case 2. The maximum decrease occurs in sea state $H_s$ = 1.5 m and $T_p$ = 10 s with −26.6 kW. This behaviour suggests that, in sea states with long peak periods and small wave heights, a considerable change of transmission ratio between one time window and the next is necessary (as applied to Case 4). Case 4 shows greater improvements compared to Case 3 throughout the power matrix and the increase in power extraction depends primarily on $H_s$ (Fig. 6b) with a maximum increase of 0.46 MW at $T_p$ = 11 s. Compared to Case 3, this degree of adaptability increases power output at lower significant wave heights, while it roughly doubles the power output increase at greater significant wave heights ($H_s$ > 4 m). In specific peak periods, it is noticeable that the difference in power output increases relative to the neighbouring sea states: for Case 3 this occurs at $T_p$ = 8 s, while for Case 4 the effect can be seen at $T_p$ = 7 s and 11 s. This could be cause by constructive or destructive hydrodynamic effects within the WEC array elements. The role of constructive (or destructive) interference on the total power extraction of the array will be further investigated in future work by looking at the behaviour of each array element, as was done in previous work for regular waves [29].

### 4.2. Capture width ratio

The CWR puts the absolute power extraction of the device into context by comparing it to the available power from the incoming wave; a large dense WEC array such as the OG3.0 with a total width of 52 m (equal to the array diameter) has a large incoming wave range it can absorb power from. It is important to consider this metric, as the power matrices might give a distorted impression of the sea states in which the WEC array performs best: higher total power output does not necessarily mean that the device operates more efficiently at the location where it has been deployed.

Fig. 7 shows the range of CWR for the four different cases and underlines the findings analysed in the previous section. In Case 1, the OG-WEC captures less than 10% of the incoming wave energy in sea states with $H_s$ < 3 m and in all sea states with $T_p$ > 12 s. In Case 2, the CWR increases at lower $H_s$ between the range of $T_p$ = 6–12 s but does not surpass 0.2. As mentioned considering Fig. 6, Case 3 does not improve the device's performance for long peak periods (11–14 s) and low significant wave heights where the CWR does not surpass 0.1. For Case 3, it is possible to identify slightly higher CWR around $T_p$ = 7 s; a maximum of 0.25 is reached at higher $H_s$ and, at the same time, the range in which this CWR is reached is broadened. Case 4 reaches a CWR of up to 0.3 at $T_p$ = 6 s and extends the CWR especially at higher $H_s$. By analysing the CWR, it becomes clearer that, at low $H_s$ for all cases, the performance is mainly influenced by $H_s$, where delimitation lines are roughly horizontal (less so for Case 1). For higher $H_s$ (from 2 m onward), $T_p$ has most influence on the CWR.

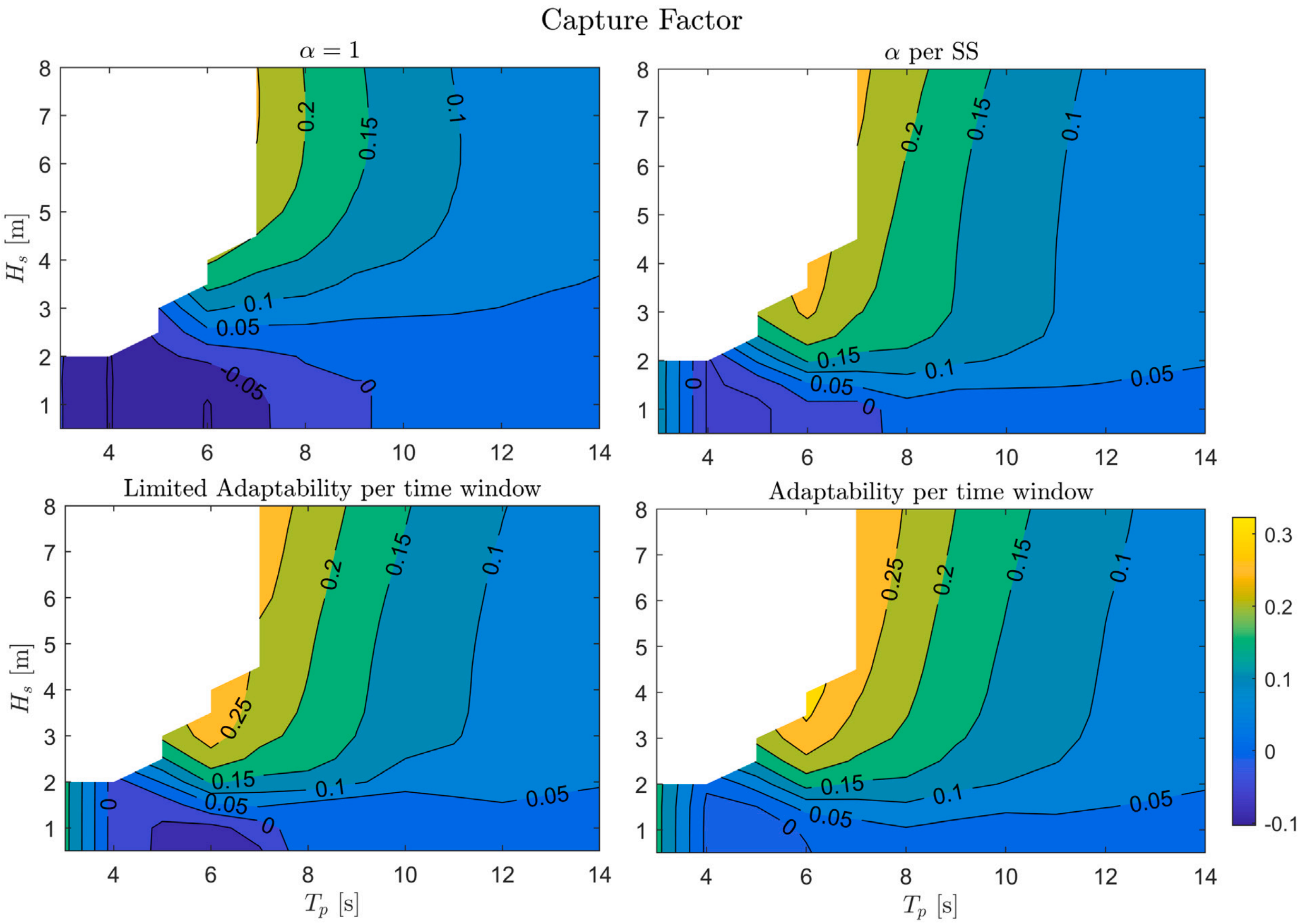

**Fig. 7.** Capture width ratio for all adaptability level cases.

**Table 3**
AEP of all cases.

| Case | AEP [GWh] |
|---|---|
| Available | 8.63 |
| Case 1 | 0.54 |
| Case 2 | 1.07 |
| Case 3 | 1.10 |
| Case 4 | 1.20 |

### *4.3. Annual energy production*

Placing the OG-WEC with different degrees of adaptability in a specific location (Section 3) allows for the calculation of the AEP. From the scatter diagram depicted in Fig. 3, the occurrence of each sea state in hours per year is multiplied with the power matrices of Section 4.1. The AEP can be depicted in matrix form as shown in Fig. 8 and summed to give the total annual energy produced by the device ( Table 3).

The most occurring and most energetic sea states, as depicted in Fig. 3, yield slightly higher CWR for Cases 3 and 4. Cases 1, 2 and 3 show noticeable negative power extraction (and AEP) in the lower part of the scatter diagram. These findings lead to the differences in total energy produced by the different cases in this location that can be seen in Fig. 8. Table 3 shows that, without any adaptability, the device only produces 0.54 GWh which leads to an average CWR of 0.06, predominantly influenced by the negative values in the range of $H_s = 0.5$–$2$ m and $T_p = 3$–$9$ s. Again, it should be noted that these values are not physically meaningful as the actual device would not consume energy, but they highlight the fact that the device and PTO system without adaptability cannot operate successfully at these sea states and, therefore, the chosen location. As the area in which negative energy can be found is significantly smaller for Case 2 ($H_s = 0.5$–$1.5$ m and $T_p = 4$–$7$ s) the total AEP already shows a significant increase reaching more than 1.07 GWh and an average CWR of 0.12. Case 3 did not show significant improvements (and even showed worse performance in lower $H_s$) from the power matrix results; its efficiency of energy extraction is shown to be better in sea states with higher $H_s$ which leads to a slightly higher total AEP of more than 1.1 GWh. Comparing Case 4 with Case 2 in this location shows an increase of 12% in AEP with the total being larger than 1.2 GWh. These slight differences are due to the CWR and performance of the WEC array being improved especially in sea states with higher $H_s$ that do not occur as frequently in the chosen location.

It is necessary to note that this research does not include limitations to the floaters' elevation, velocity or to the forces acting on them, as taken into consideration in other research works, such as De Backer et al. [10]. Thus, it should be acknowledged that the presented findings of power output and performance constitute the best case scenarios for the adaptability levels investigated in this work. On the other hand, we expect that optimising the transmission ratio per element within the array will show further increases in power extraction. Future work, therefore, will focus on including mechanical interactions between floaters, as well as optimising adaptability within the array to yield more accurate predictions of power extraction and performance.

### *4.4. Transmission ratio*

For Cases 2, 3 and 4, it is possible to analyse the transmission ratios chosen to maximise the power extraction of the WEC array in the range of sea states defined previously, and depict it in matrix format (Fig. 9);

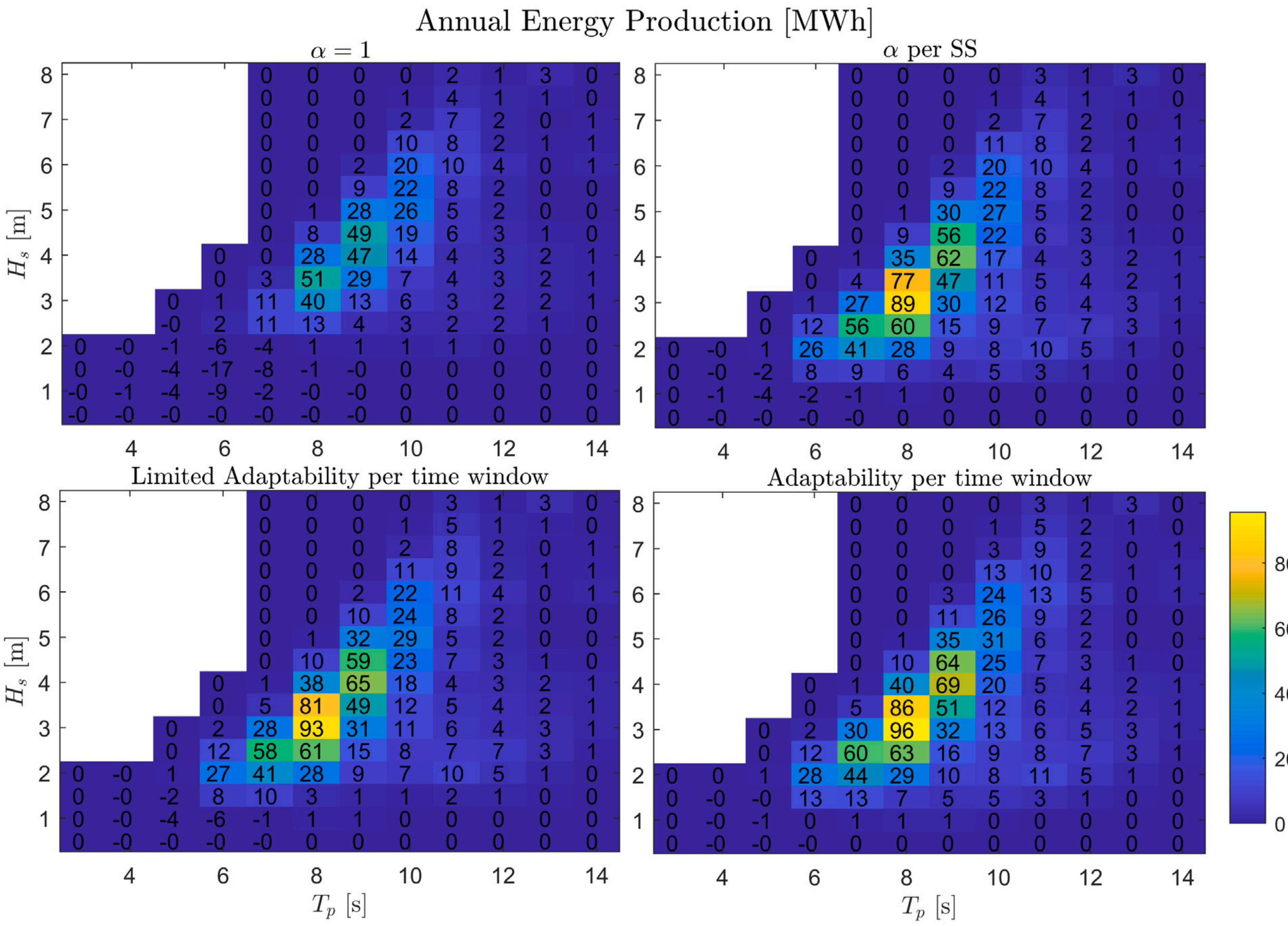

**Fig. 8.** Annual Energy Production for all adaptability level cases.

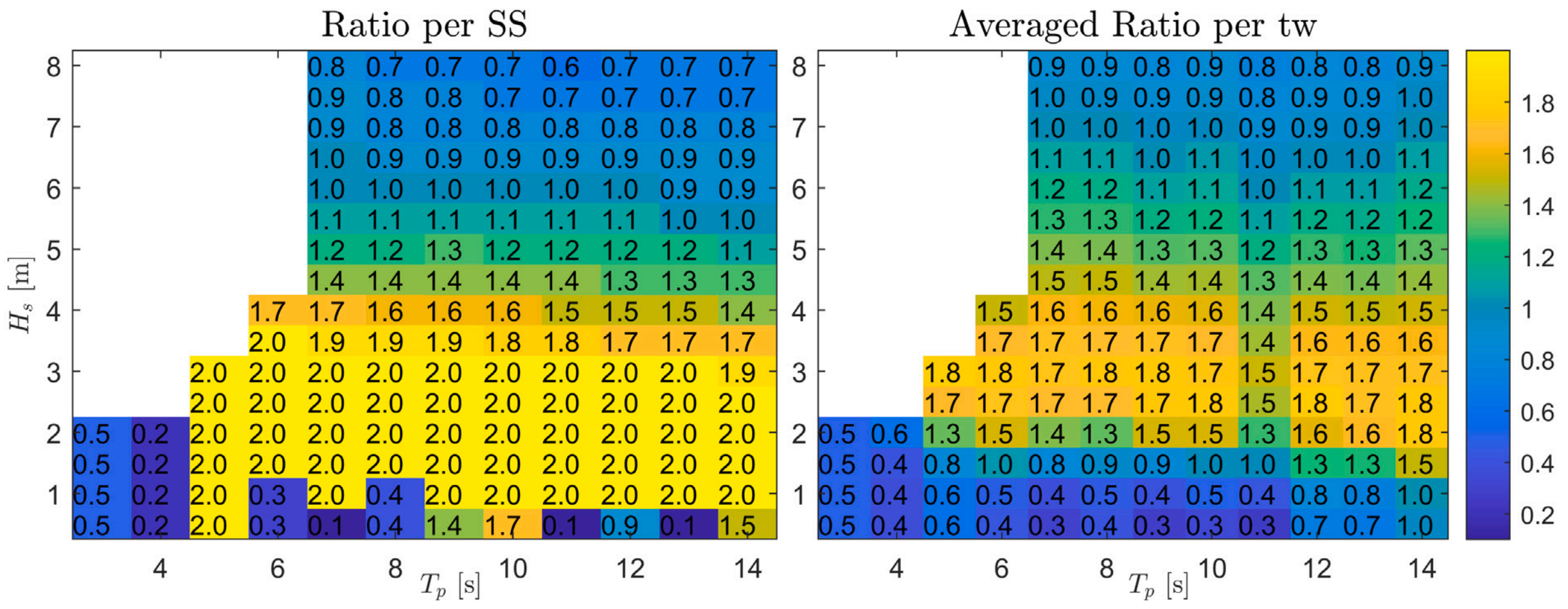

**Fig. 9.** Transmission ratio in two adaptability levels: (a) constant over the entire sea state; (b) optimised for each time window without restriction and averaged over the entire sea state.

studying this parameter can clarify what its influence is on the power extraction and how it should be tuned to maximise its benefits. The range of transmission ratio $\alpha$ needs to be critically assessed in order to draw conclusions on its applicability and significance. For Case 2, the value of $\alpha$ that maximises the power extraction is shown, while the average value of $\alpha$ over the entire sea state simulation is depicted for Case 4 in Fig. 9. On the basis of these results, the choice of $\alpha$ in specific sea states can be analysed more thoroughly.

In Case 2, in the sea states with higher energy potential (higher $H_s$ and $T_p$), the optimal transmission ratio is within the predefined transmission ratio range of 0.1–2, ensuring that the device is designed to perform best in those sea states (Fig. 9a). It can be observed that, in these sea states (as shown with the example of sea state $H_s$ = 6 m, $T_p$ = 12 s in Fig. 10a), the maximum value of extracted power is achieved within the predefined transmission ratio range. Other sea states such as the one shown for $H_s$ = 2 m, $T_p$ = 8 s (Fig. 10b), however, appear to reach maximum power output beyond the selected transmission ratio range ($\alpha$ = 2.0 in Fig. 9a). Hence, it is likely that extending the range with $\alpha$ > 2.0 will lead to a higher power extraction. In Fig. 10a and b, the power extracted reaches values close to 0 at specific transmission ratio values. This might be caused by a combination of factors influencing the power extraction of a dense WEC array such as destructive interaction effects at specific wave periods, anti-resonant behaviour determined by the geometry of the floaters, and tuning of the PTO system that causes the PTO to remain stationary leading to sticking behaviour [29,36]. Further research comparing the results to single floater performance will be able to narrow down the cause of this behaviour and quantify the contribution of each of the factors mentioned above.

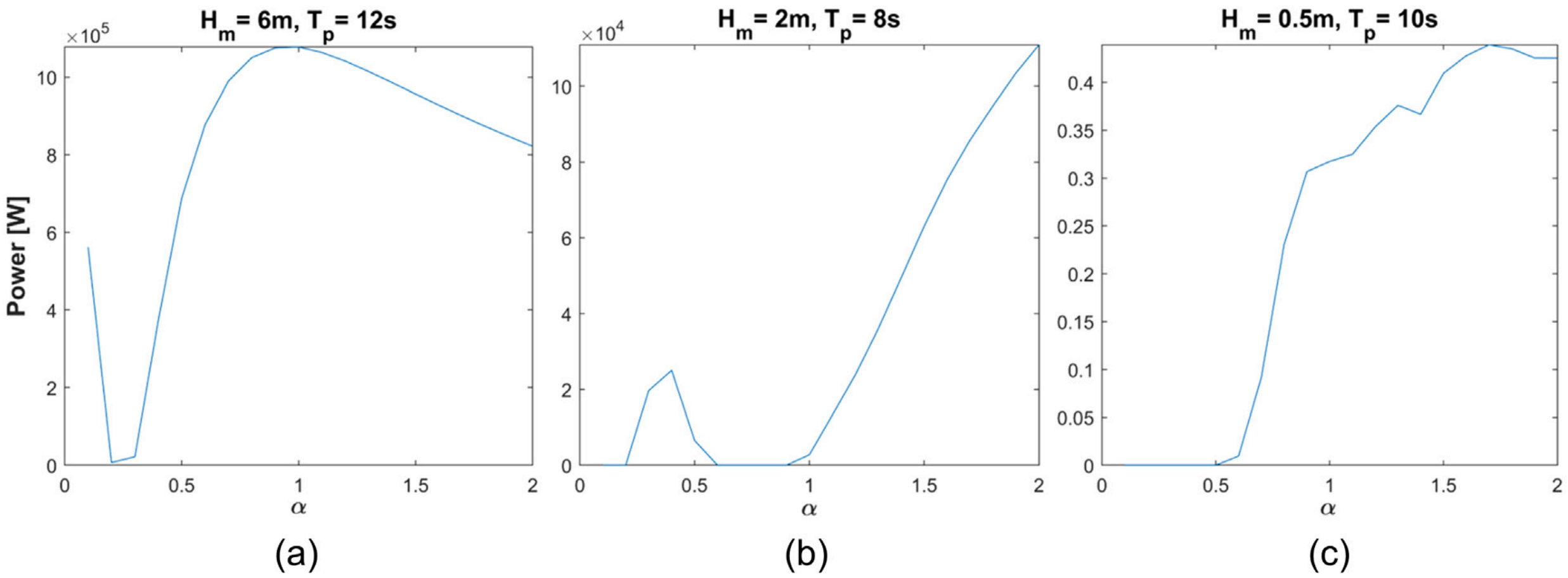

**Fig. 10.** Effect of the transmission ratio (as described in Case 2) on the average power extracted by the WEC array in specific sea states.

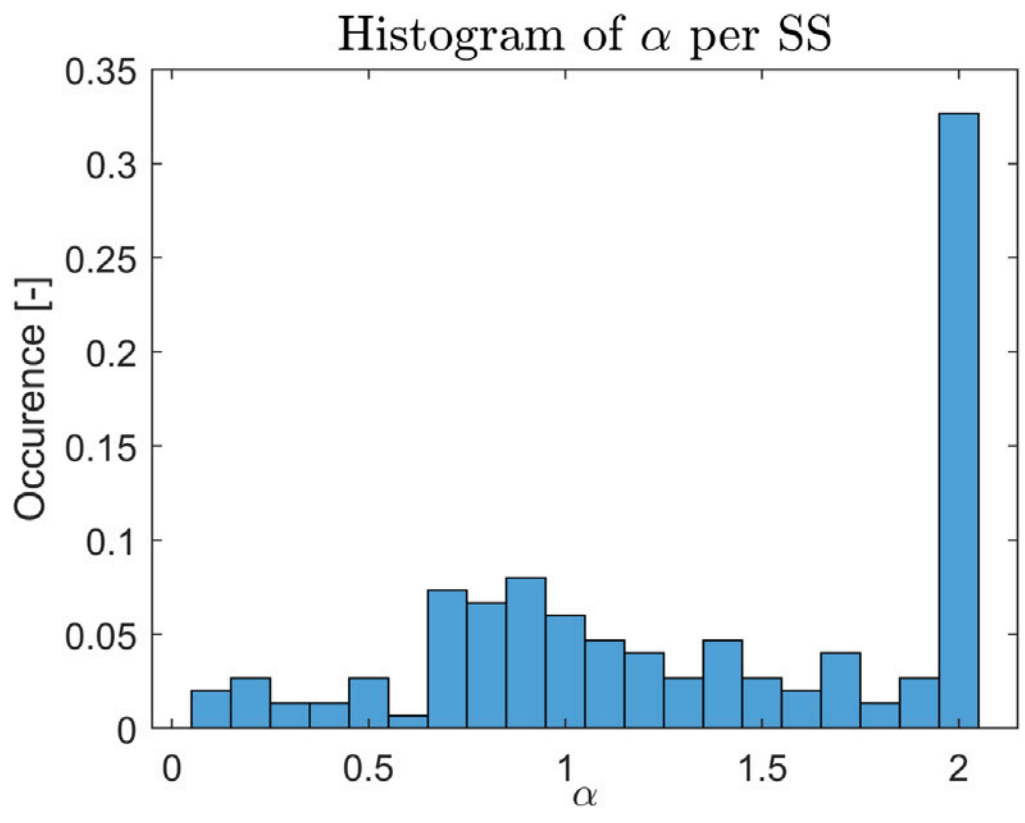

**Fig. 11.** Histogram of $\alpha$ chosen to maximise the power output of the WEC array in Case 2 for the total power matrix.

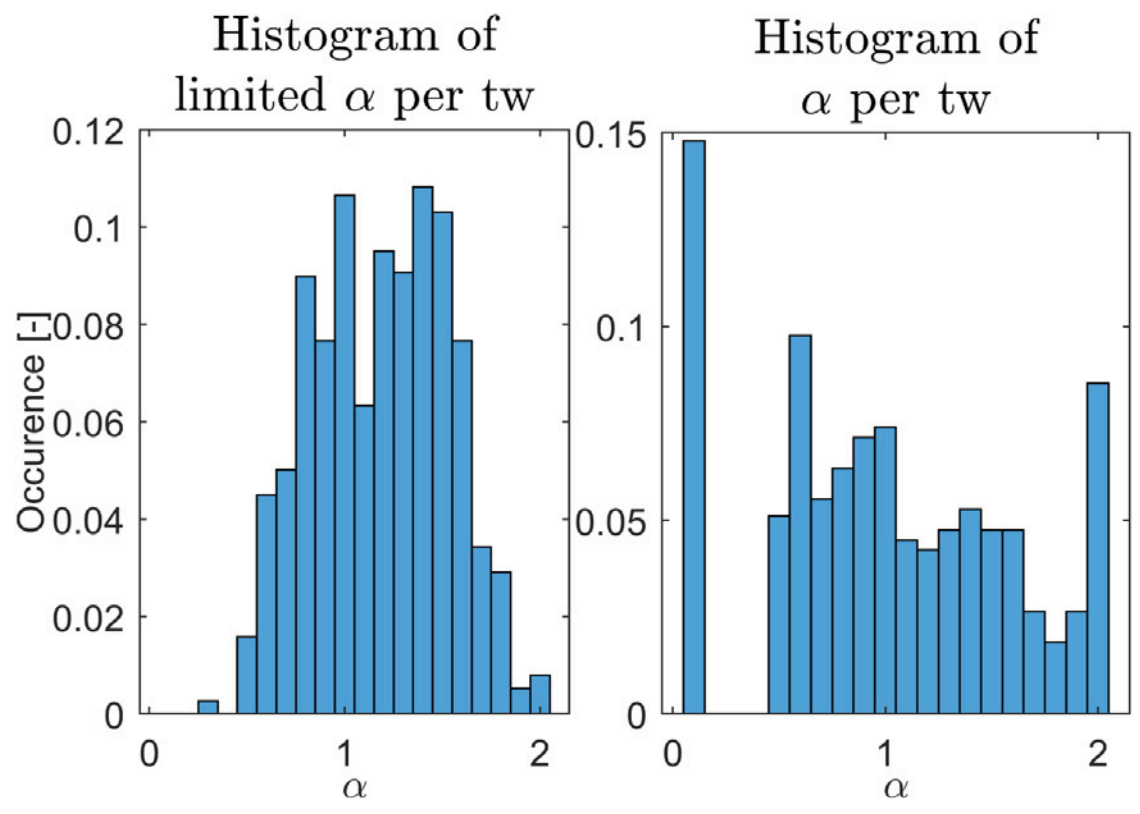

**Fig. 12.** Histogram of $\alpha$ chosen in sea state $H_s$ = 6 m and $T_p$ = 11 s for Case 3 (left) and Case 4 (right).

When reaching sea states in which the optimal transmission ratio is far outside the predefined range in Fig. 9, it is noticeable that the change of transmission ratio does not influence the performance of the WEC array significantly (as represented in Fig. 10c). Observing this behaviour in low $H_s$ and short $T_p$ in Fig. 9a suggests that, in order to optimise the WEC array's performance for those sea states, the initial design of the device would have to be adapted (e.g. the diameters of the PTO pistons, the piping length and the water depth). From the matrix in Fig. 9a it is also noticeable that the transmission ratio for Case 2 does not reach the lower bound of its range in these sea states. The histogram of all values of $\alpha$ chosen to maximise the power extraction in the total power matrix for Case 2 (Fig. 11) shows that $\alpha = 2$ yields the best results for more than 33% of the cases. In connection to the results of the power matrix (Fig. 4a), this indicates that the defined transmission ratio range and, hence, the design is not suitable for those sea states and would need to be adapted at the WEC array design parameter level (e.g. floater or PTO) to improve the performance.

With increasing level of adaptability (Case 4), the average transmission ratio for each sea state decreases for lower $H_s$ and increases for higher $H_s$ (Fig. 9b): for $T_p = 6$ s and 11 s, the average transmission ratio is appreciably smaller compared with the sea states outside of these $T_p$ values. In connection to the absolute difference in power extracted by Case 4 compared to Case 2 in Fig. 6b, the sea states in which the power difference is higher occur at the same range of peak periods. The difference in transmission ratio between Cases 2 and 4 is moderate but it is visible that the transmission ratio change for Case 4 is influenced more by $T_p$.

For Cases 3 and 4, the histograms of chosen $\alpha$ are presented to highlight the function and necessity of the transmission ratio in adapting the WEC array to a range of sea states. These show how $\alpha$ is chosen within the simulation of the sea state when maximising the power extraction for each time window (with and without restriction). Three sea states are selected to show how the range of $\alpha$ was used: in the first sea state, an almost uniform distribution can be seen, while in the second and third, the occurrence of alpha is skewed to the left or to the right of the range. Ideally, the WEC array would be tuned such that the transmission ratio remains in the middle of the range in the most frequently occurring and energetic sea states of a location; in this way, the limits of the range can be used to adapt the array performance for energetically higher and lower sea states. The histogram of sea state $H_s$ = 6 m and $T_p$ = 11 s in Fig. 12 for Case 3 shows a normal distribution of $\alpha$ around the centre of the range, while Case 4 is closer to a uniform distribution. Similarly to what can be observed in Fig. 10 a and b, certain transmission ratios were not selected in the analysed sea state for Case 4. This could be the effect of destructive interference within the WEC array, anti-resonant behaviour of single floaters and the PTO force being wrongly tuned, as mentioned above. Future research focusing on the performance of single floaters within the WEC array is expected to give more details about this behaviour.

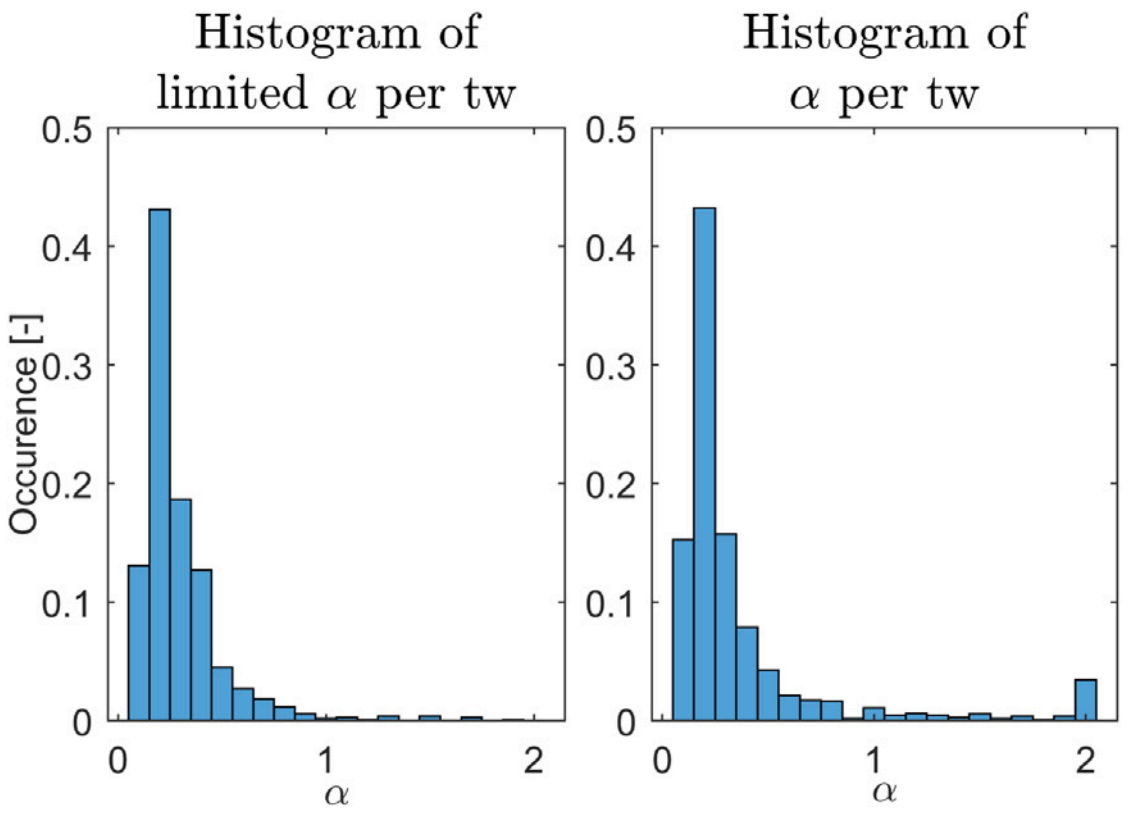

**Fig. 13.** Histogram of $\alpha$ chosen in sea state $H_s = 1$ m and $T_p = 4$ s for Case 3 (left) and Case 4 (right).

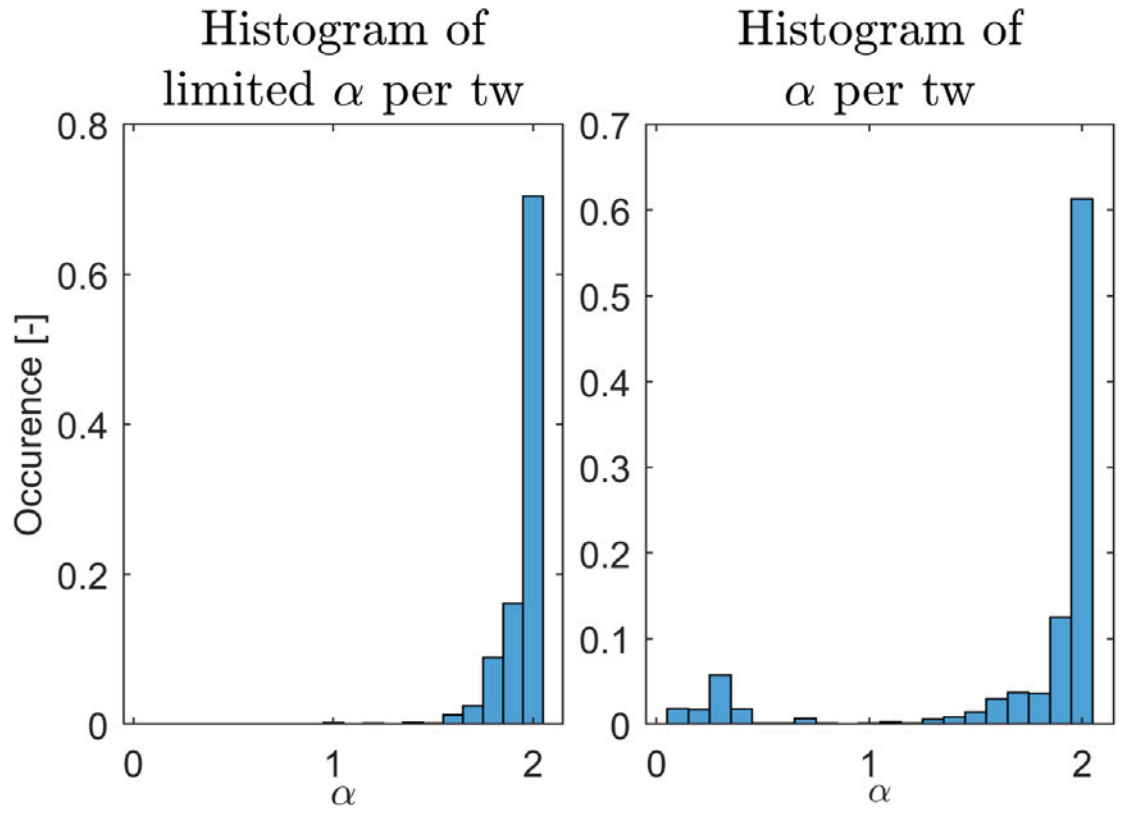

**Fig. 14.** Histogram of $\alpha$ chosen in sea state $H_s = 2.5$ m and $T_p = 6$ s for Case 3 (left) and Case 4 (right).

In the sea state plotted in Fig. 13, both adaptability cases show a Rayleigh distribution skewed to the left of the transmission ratio range, with Case 4 showing a slight increase at the maximum value of $\alpha$. The two cases show more similar $\alpha$ selection compared to Fig. 12. More than 40% of the time, $\alpha$ is selected to be 0.2; this might suggest that extending the range to include smaller values or redesigning the initial PTO set-up might increase the performance efficiency at this and similar sea states. The results of Case 4 in connection with Fig. 10c, might also suggest that increasing the transmission ratio range at the higher limit ($alpha > 2.0$) might improve performance; this is further supported by Fig. 7 for Case 3 and 4 that shows negative or close to 0 values of CWR for this sea state.

To further highlight the importance of selecting the range of $\alpha$ properly, sea state $H_s = 2.5$ m and $T_p = 6$ s in Fig. 14 shows the opposite behaviour than Fig. 13: both adaptability cases show a strong skew to the high end of the transmission ratio range, suggesting that extending the range on that side might lead to higher power output. As in the previously analysed sea state, also in this case a higher adaptability (Case 4) leads to a small increase in selection of ranges in the opposite end of the range. This suggests that the ability to switch from high to low $\alpha$ within a sea state increases the performance.

## 5. Conclusion

In the present work, the adaptability of the PTO system within a WEC array is analysed. This adaptability is approximated by a transmission ratio between the output and input force (into the PTO versus the output from the floater), and four cases are defined to study different degrees of adaptability. On the basis of the OG-WEC case, simulations are performed to calculate the power extraction and transmission behaviour in a range of sea states to create comparable power matrices. These findings are further put into context by the CWR, and the AEP for a defined location in the North Sea. By comparing the transmission ratio behaviour and power output of the four cases, the following conclusions can be drawn:

- Including adaptability in the PTO system of a WEC (array) device increases the power output, with more adaptability leading to more power output.
- The range of adaptability must be carefully selected in order to include the necessary settings to maximise the efficiency of the WEC array in the desired sea states, while also keeping in mind the physical constraints of the selected PTO system.
- When the adaptability needs to be limited, a well defined strategy on how to tune the device is necessary; otherwise, adaptability may end up decreasing the efficiency at certain sea states.
- In general, the adaptability will increase the power output, but the way in which the adaptability range is chosen needs to be defined through the physical parameters of the PTOs powered by the WEC array.
- Adaptability might enhance constructive effects within dense WEC arrays in specific sea states.

This research analyses the power extraction of the OG-WEC array without including constraints to floaters' displacements or the forces acting on them; including specific limitations is expected to yield more realistic findings and may alter the influence of adaptability. The performance of each floater within the dense WEC array with different adaptability ratios $\alpha$ could give more insight into the effects of adaptability on the interactions between the floaters in the array. Furthermore, the advantages of adaptability and WEC array configuration have not been fully exploited yet; as previous research suggests [10,37], it is most valuable to tune each floater individually within a WEC array. With simulation models making it possible to include highly complex and interrelated WEC arrays while also decreasing computational efforts, future research will study an adaptable transmission ratio applied separately to each floater-PTO pair within a WEC array. This could lead to further improvements in power extraction. Finally, a physical PTO system with adaptability needs to be studied in order to compare the advantages of a continuous, theoretical adaptability parameter ($\alpha$) to a discrete adaptable settings.

## CRediT authorship contribution statement

**Alva Bechlenberg:** Conceptualization, Data curation, Investigation, Methodology, Validation, Visualization, Writing – original draft. **Yanji Wei:** Conceptualization, Methodology, Software, Validation, Writing – review & editing. **Bayu Jayawardhana:** Supervision, Writing – review & editing. **Antonis I. Vakis:** Conceptualization, Supervision, Writing – review & editing.

## Declaration of competing interest

One or more of the authors of this paper have disclosed potential or pertinent conflicts of interest, which may include receipt of payment, either direct or indirect, institutional support, or association with an entity in the biomedical field which may be perceived to have potential conflict of interest with this work. For full disclosure statements refer to https://doi.org/10.1016/j.renene.2023.04.076.A.I. Vakis reports a relationship with Ocean Grazer B.V. that includes: board membership. B. Jayawardhana reports a relationship with Ocean Grazer B.V. that includes: board membership. A. I. Vakis has patent #WO2019117711 issued to Rijksuniversiteit Groningen. B. Jayawardhana has patent #WO2019117711 issued to Rijksuniversiteit Groningen.

## Acknowledgements

Alva Bechlenberg is thankful to Giulia Cervelli for helpful discussions on the ERA5 dataset used in the manuscript. The authors thank the Centre for Information Technology of the University of Groningen for their support and for providing access to the Peregrine high performance computing cluster.